\documentclass[nopreprintline,12pt]{elsarticle}

\usepackage{amssymb}

\begin{document}

\begin{frontmatter}



\title{FiN: A Smart Grid and \\Power Line Communication Dataset}


\author[inst1]{Christoph Balada}\ead{christoph.balada@dfki.de}
\author[inst1]{Sheraz Ahmed}\ead{sheraz.ahmed@dfki.de}
\author[inst1]{Andreas Dengel}\ead{andreas.dengel@dfki.de}

\affiliation[inst1]{organization={German Research Center for Artificial Intelligence (DFKI)},
            city={Kaiserslautern},
            postcode={67663}, 
            country={Germany}}

\author[inst2]{Max Bondorf}\ead{bondorf@uni-wuppertal.de}
\author[inst2]{Nikolai Hopfer}\ead{nikolai.hopfer@uni-wuppertal.de}
\author[inst2]{Markus Zdrallek}\ead{zdrallek@uni-wuppertal.de}

\affiliation[inst2]{organization={University of Wuppertal},
            city={Wuppertal},
            postcode={42119}, 
            country={Germany}}

\begin{abstract}
The increasing complexity of low-voltage networks poses a growing challenge for the reliable and fail-safe operation of electricity grids. 
The reasons for this include an increasingly decentralized energy generation (photovoltaic systems, wind power, etc.) and the emergence of new types of consumers (e-mobility, domestic electricity storage, etc.).
At the same time, the low-voltage grid is largely unmonitored and local power failures are sometimes hard to detect. 
To overcome this, power line communication (PLC) has emerged as a potential solution for reliable monitoring of the low-voltage grid.
In addition to establishing a communication infrastructure, PLC also offers the possibility of evaluating the cables themselves, as well as the connection quality between individual cable distributors based on their Signal-to-Noise Ratio (SNR).
The roll-out of a large-scale PLC infrastructure therefore not only ensures communication, but also introduces a tool for monitoring the entire network.
To evaluate the potential of this data, we installed 38 PLC modems in three different areas of a German city with a population of about 150,000 as part of the Fühler-im-Netz (FiN) project.
Over a period of 22 months, an SNR spectrum of each connection between adjacent PLC modems was generated every quarter of an hour.
The availability of this real-world PLC data opens up new possibilities to react to the increasingly complex challenges in future smart grids.
This paper provides a detailed analysis of the data generation and describes how the data was collected during normal operation of the electricity grid. 
In addition, we present common anomalies, effects, and trends that could be observed in the PLC data at daily, weekly, or seasonal levels.
Finally, we discuss potential use cases and the remote inspection of a cable section is highlighted as an example.
\end{abstract}




\end{frontmatter}


\section{Introduction}
\label{sec:introduction}
Despite of the advances in IoT and increasingly cheaper sensory hardware, low-voltage grids remain largely unmonitored, as this is uneconomical due to their high complexity \cite{miller2015status}.
However, the aspect of comprehensive monitoring has become more important due to the growing challenges arising from the transition to a higher share of renewable energy and more complex energy consumers.
On the way to meet these challenges, power line communication (PLC) has proven to be an efficient way to connect households, cable distributors and grid operators \cite{sendin2014strategies,dede2015smart}.
By providing such a mesh network, interconnecting all stakeholders, PLC is a good candidate to back the operation of many different smart devices like chargers, meters, photovoltaic systems (PV), etc. 
Beside of this potential as backbone for smart grid devices, we want to shed light on another set of applications, that could also be realized by using data from such a PLC network.
While the communication capability of PLC networks is already receiving attention in research, the exploration of the benefits of the data collected in PLC networks remains largely unaddressed.
By attaching sensors to PLC nodes, it is possible to collect a wide range of different types of data. 
In this paper, however, we want to focus on the Signal-to-Noise Ratio (SNR) measured by the PLC modems during their normal operation and without additional sensors. 
\begin{figure}[t!]
    \centering
    \includegraphics[width=0.7\textwidth,clip]{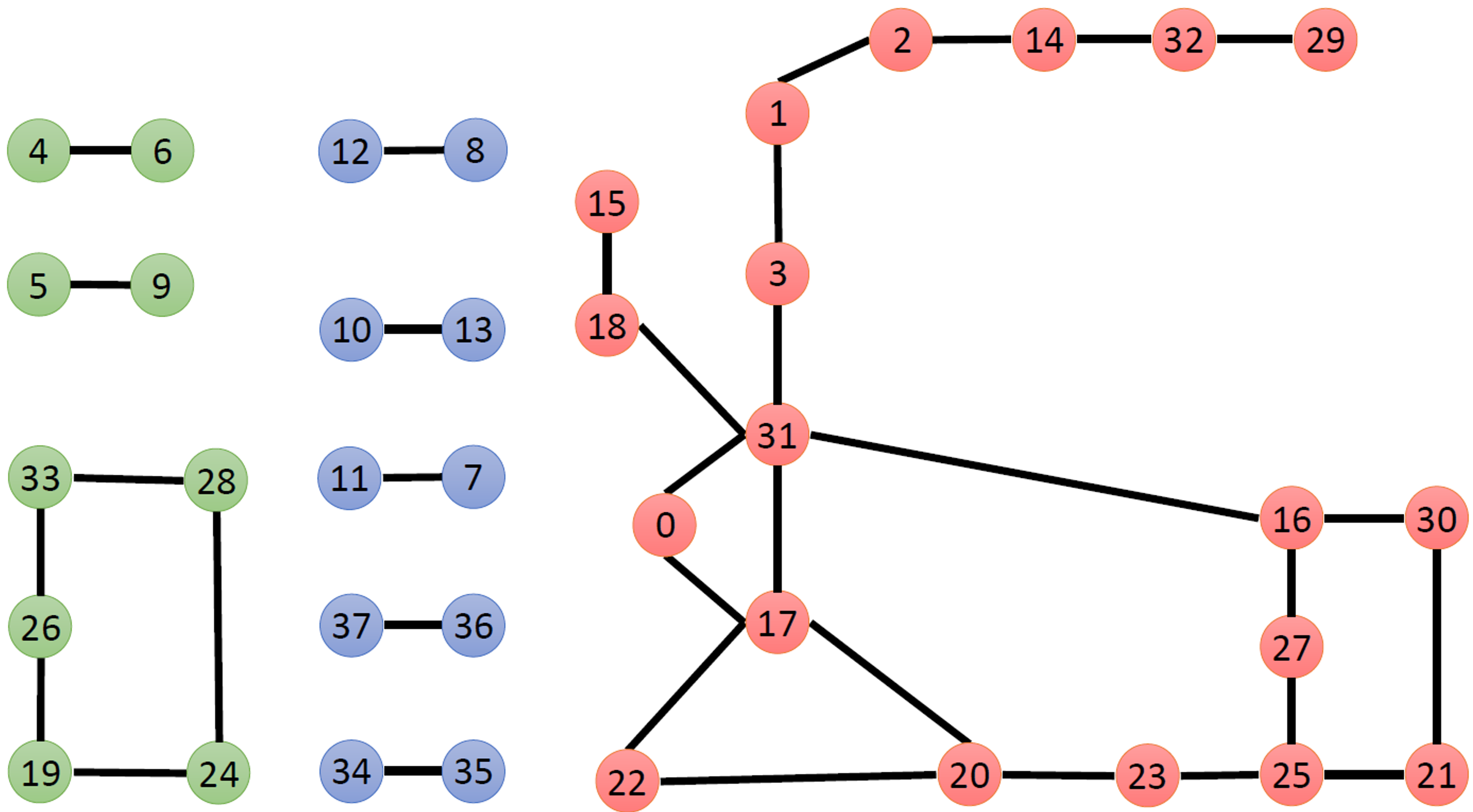}
    \caption{Grid graph of all three locations. Location 1 in green, location 2 in blue, location 3 in red. Topology that is formed by the PLC infrastructure. The numbers represent the node ID, which is also used within the dataset. An example of a visible neighborhood would be when node 22 can communicate with node 31. In this case, both are directly connected from the communication point of view, but there is no direct electrical connection because node 17 separates both nodes.}
    \label{fig:graph}
\end{figure}
\begin{figure*}[h!]
    \centering
    \includegraphics[width=\textwidth,trim={0cm 0cm 0cm 0cm},clip]{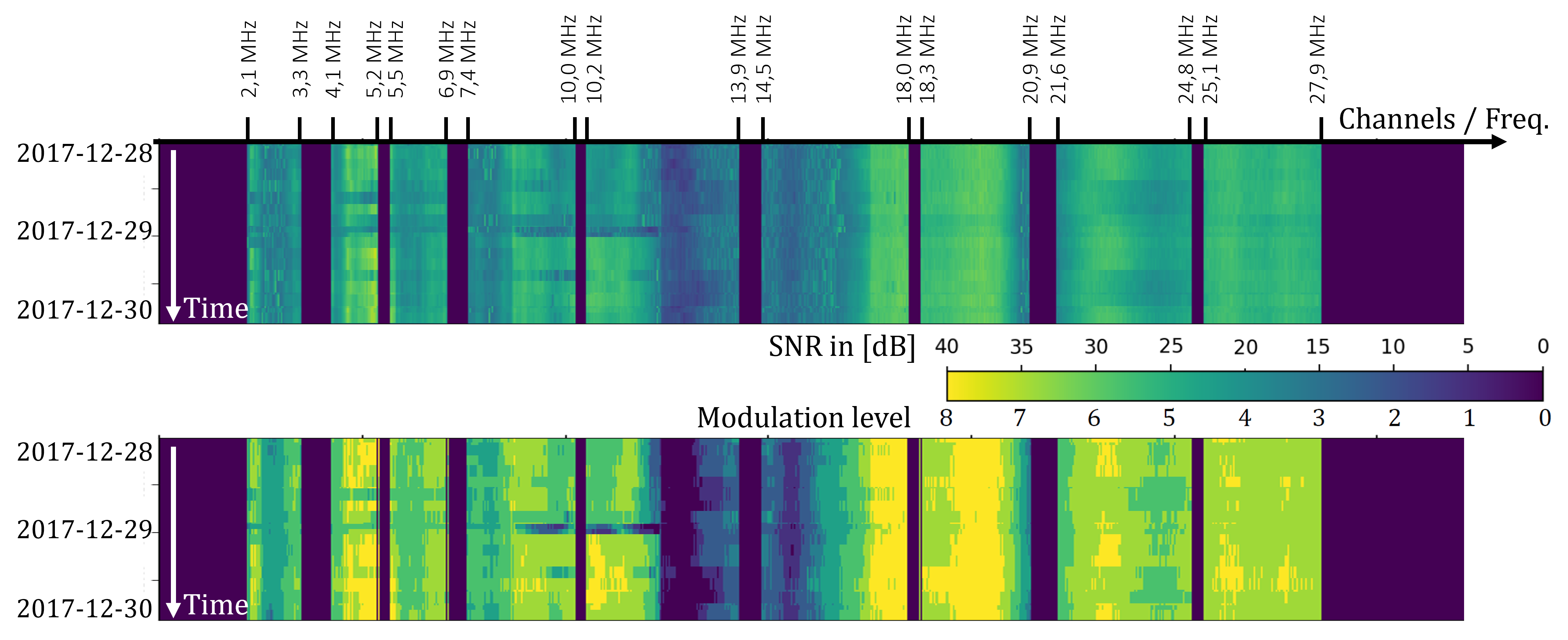}
    \caption{Top: SNR spectrum for a single connection. The spectrum illustrates the SNR spectrum for 192 assessments that correspond to two days. Dead spaces in the spectrum are due to channels or frequencies that are not cleared for use in PLC communication. Of about 1500 channels, 917 remain useful channels. Bottom: Tonemap for a single connection. The tonemap illustrates the modulation level for 192 assessments.}
    \label{fig:snr_tm_two_days}
\end{figure*}
One reason why this data is largely unconsidered in research is not least because data from the practical use of PLC networks has not yet been sufficiently available to the research community.
In this work, we want to address this shortcoming by presenting the FiN dataset and potential use cases. 
Within FiN 38 PLC modems were distributed in three different areas of a German city with population of about 150,000.
Those 38 modems yield 68 1-to-1 connections and collected data over around two years.
During this time every quarter of an hour an SNR assessment as described in section \ref{sec:snr_assessment} was performed.
Thus over 3.7 million data points were collected during the project period. 
Along with this SNR data, we publish an extensive collection of metadata on the cables in the network. 
To underline the benefits of the data, we also highlight a number of different applications.

\subsection{Main contributions}
We present the FiN dataset holding 3.7 million data points, where each data point represents the SNR profile of a PLC connection between two cable distributors in the low- and mid-voltage grid. 
Furthermore, we provide metadata on weather conditions and cable characteristics.
All the data was recorded during the real-life usage of the PLC nodes. 
The signal quality is measured by assessing the SNR on 917 channels with different frequencies between 2 and 30 MHz. 
The data was collected over a period of 22 months and involves 38 different PLC nodes inside the cable distributors.
FiN dataset aims on a wide range of different potential applications:
\begin{itemize}
    \item Anomaly detection (e.g. discovery of noise sources) 
    \item Grid monitoring (e.g. detection of power outages, localization of cable breaks)
    \item Asset management (e.g. lifetime estimation of cables, meta data estimation)
    \item Additional security layer (e.g. detection of attacks against the grid infrastructure, manipulations)
    \item Misc (e.g. Cloudiness estimation through photovoltaic systems) 
\end{itemize}

This broad range of applications highlights the potential of this data, but on the same side also the complexity of evaluating every use case in detail.

In this paper, we propose a novel dataset consisting of over 3.7 million real-world PLC measurements (S-\ref{sec:properties}). 
The data acquisition process is described in detail in S-\ref{sec:locs}.
In S-\ref{sec:patterns}, we provide a first analysis of trends and different types of interference in the FiN dataset and S-\ref{sec:apps} discusses the aforementioned fields of application in detail. 
Our analysis highlights the uniqueness of the proposed PLC dataset, and it’s relevance for a variety of future applications. 
This paper should raise awareness and encourage the research community to further investigate the various possibilities of FiN and similar datasets.

\section{Related Work} \label{sec:relatedWork}
Electricity grids are an essential part of the economic and personal lives of most people on earth.
However, the electricity grid is presently undergoing one of the biggest transformations since its invention \cite{miller2015status}.
Climate changes are unquestionable and with rising urgency for changes in the way we produce, consume, store and transmit electrical power, new approaches for our electrical grid become inevitable.
On the other hand, digitalisation and automation are opening up new ways of dealing with these changes.
To address both problems many different aspects of future grid systems, often called smart grids, need to be rethought. 

A big number of research works concern especially forecasting of future demand in electrical power to optimize generation and demand \cite{rolnick2019tackling,hippert2001neural,soliman2010electrical,hong2016probabilistic,alzahrani2017solar,sun2018solar,wan2015photovoltaic,mathe2019pvnet}.
Since it is a complex task to store significant amount of electrical power, it is the goal to achieve an equilibrium between generation and demand. 
\cite{candanedo2017data,UNdata,pjmdata,tsanas2012accurate,usdata} show different datasets and works concerning electrical load forecasting and address different scenarios on hourly or daily basis, for households, big buildings, residential buildings, cities or whole countries.
Especially due to financial interests, the topic of consumption forecasting receives a lot of attention.
In the context of the transformation towards a smart grid, however, this area only covers one aspect. 
Topics such as asset monitoring, grid automation or grid security benefit comparatively little from this financial interest.
In contrast, the FiN dataset attempts to address these less highlighted use cases in particular and it's SNR data from practical application adds another important building block that will support research.
In the following, an overview of other possible fields of application of SNR data is given, even though these were mainly carried out on simulated data and were not evaluated in practice.
    
\begin{itemize}
    \item \textbf{Asset monitoring} deals with the monitoring and assessment of the grid assets.
    This includes, for example, the recording of the ageing process and partial discharges \cite{hopfer2017new,hopfer2017identification,rezaei2018new,forstel2017grid} or the detection of events such as a fuse failure \cite{bondorf2021broadband}.
    As \cite{hopfer2017new,hopfer2017identification,rezaei2018new,bondorf2021broadband} show, SNR data is also suitable for this area. 
    Section \ref{sec:sleeves} shows an example of how the FiN data can be used to estimate the number of connecting joints installed without manual measurements on site.
    
    \item \textbf{Security} is also of increasing importance due to the increasing digitalisation of the electricity grid. 
    More intelligent controlling also opens up more routes of attacking the electricity grid.
    Additional devices that supervise the correct operation of the grid and that act independently of the normal grid operation have been successfully tested in \cite{he2019power}. 
    The PLC infrastructure can be used analogously for decentralised, secure and independent monitoring of the network status \cite{jin2018efficient}.
    The FiN data can be used to enable research on real data.

    \item \textbf{Grid stability and automation} is a topic of major importance, especially due to the increasing complexity of the electricity grid. 
    Many different works \cite{schafer2016taming,nord2015prof,arzamasov2018towards,kamps2020reliability,ludwig2019multi} deal with different aspects and techniques to ensure stability or automation.
    However, most of them have in common that they are significantly based on grid parameters, such as voltage or frequency.
    The FiN dataset, together with the SNR data, provides an additional new foundation of data, which also opens up new scope for research in this area.
    
\end{itemize}

\section{Properties of FiN}\label{sec:properties}
FiN consists of 3.7 million SNR assessments collected from 38 PLC nodes that form 34 connections, since each connection is measured in both directions, the dataset shows overall 68 connections.
A highlight feature of FiN is that the data was collected over a period of about 2 years and recorded entirely in a real-life environment. 
Therefore, the FiN data covers various weather conditions, seasonal effects, noise sources, trends and sudden changes on mid-voltage (mv) as well as on low-voltage (lv) level. 
The data was collected from 12th of April 2017 to 5th of February 2019 while 14 PLC modems are installed on mid-voltage level and 24 on low-voltage level. 
Apart from the SNR measurements, metadata was collected for each PLC modem.
Most important metadata are cable related data (length, type, the number of wires, cross-section, age, number of t-joints) and environmental related data (daytime, weather, characterization of the surrounding area). 
See table \ref{tab:metadata} for an overview of all available metadata information. 
This section provides a detailed overview of how the data was collected and table \ref{tab:keyfacts} summarizes the key facts.

\begin{table}[h]
    \centering
    \begin{tabular}{lc}
        Dataset key facts & \\
        \hline 
        Number of nodes & 38 \\
        Number of nodes on low voltage level & 24 \\
        Number of nodes on mid voltage level & 14 \\
        Number of 1-to-1 connections & 68 \\
        Number of data points & 3.747.610 \\
        Number of different locations & 3 \\
        Measurement frequency & 15min interval \\
        Measurement begin & April 2017 \\
        Measurement end & February 2019 \\
        Meta data & Weather and cable characteristics \\
    \end{tabular}
    \caption{A summary of the main characteristics of the dataset.}
    \label{tab:keyfacts}
\end{table}

\subsection{Device Locations}\label{sec:locs}
\begin{figure}[!b]
  \begin{center}
    \includegraphics[width=0.6\textwidth,trim={0.0cm 0.0cm 0.0cm 10.0cm},clip]{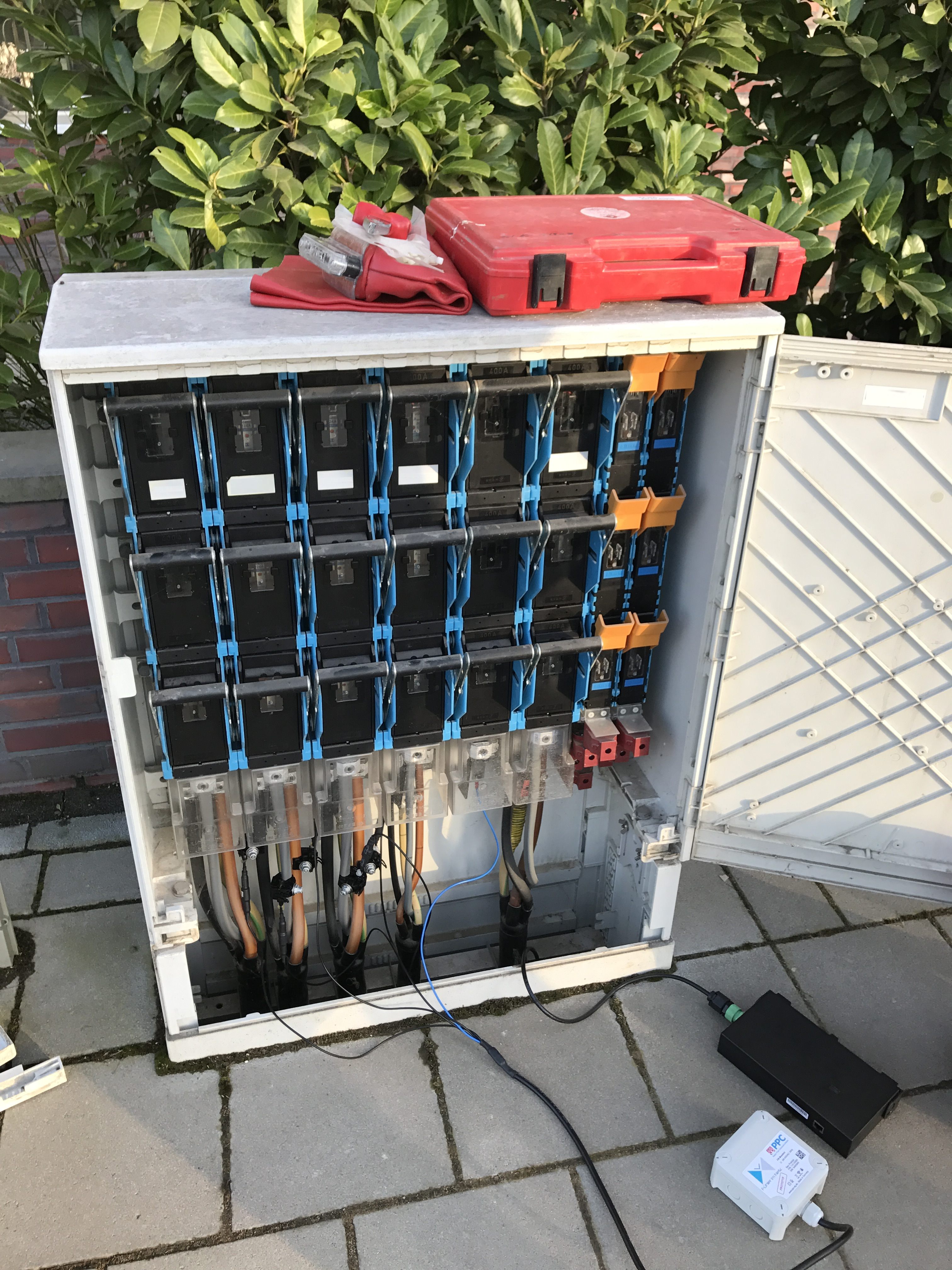}
  \end{center}
  \caption{A typical FiN installation. The PLC modem (black box) is connected to one phase and remains in the cable distribution cabinet. }
  \label{fig:finbox}
\end{figure}

Overall 38 PLC modems were installed in three different areas of a German city with around 150,000 inhabitants.
Furthermore, the PLC modems are certified under the IEEE 1901 OFDM FFT Access standard.
Figure \ref{fig:finbox} shows an example of how a FiN installation looks like.
Each location shows different types of surroundings and covers different types of consumers.
The following gives a short overview of the three locations and figure \ref{fig:graph} illustrates the network topology of all 38 PLC nodes. 

\subsubsection{Location 1 - low-voltage area}
Residential area, small houses / single family houses, only few and small photovoltaic setups, adjacent to a big office complex and smaller industry buildings with a big photovoltaic setup.
\subsubsection{Location 2 - mid-voltage area}
Residential area, mainly big houses / apartment buildings, only few photovoltaic setups, adjacent to a residential area with many small photovoltaic setups.
\subsubsection{Location 3 - low-voltage area}
Residential area, single family houses and big apartment buildings, only few and small photovoltaic setups, adjacent to a swimming pool complex and multiple schools.

\subsection{Electrical and visible neighbours}\label{sec:in_direct_neighbours}
Since all PLC modems share the cable as communication medium, they are able to talk not even to direct neighbours but also to neighbours that are multiple hops away. 
Therefore, it is necessary to distinguish between electrical neighbours, which are physically adjacent to each other and visible neighbours that can establish a PLC connection to each other.
To keep the amount of data and redundancy low, we have decided to omit visible neighbours that are no physical neighbours.
Since only electrical neighbours are shown in this dataset (see figure \ref{fig:graph}), they are simply referred to as neighbours in the following. 

\subsection{Operation time and timeouts}
Unless otherwise stated, PLC modems run 24/7. 
However even in case of a high quality connection of a PLC modem we observed that timeouts occur as part of the normal behaviour.
The reasons for timeouts are manifold and will be discussed further in section \ref{sec:patterns}. 
However, if a PLC modem crashes it automatically reboots and continues its normal operation cycle, which is started with an initial SNR assessment.
This initial assessment is one reason for measurements that occurred outside the 15-minute slots

\subsection{SNR assessment}\label{sec:snr_assessment}
Every quarter of an hour, PLC modems stop their normal communication task and perform an assessment of the connection quality by measuring the Signal-to-Noise Ratio. 
Therefore each PLC modem sends a predefined signal while its neighbors are listening for this signal. 
Based on the discrepancy between sent and received signal the SNR is computed. 
Since PLC uses a range of frequencies, the SNR is computed for 917 channels used by the modem.
The temporal progression of the SNR values over all frequencies is referred to as the SNR spectrum. 
Figure \ref{fig:snr_tm_two_days} shows an example of an SNR spectrum for 192 assessments that correspond to two days. 
Apart from the SNR values, PLC modems also capture the modulation level that was used to encode and decode the PLC signal. 
The modulation level is given in eight discrete steps and indicates the number of bits that are transmitted per signal word. 
While modulation level is captured in receiving and transmitting direction, the SNR values are only computed for the received signal. 
Similar to the SNR spectrum, the modulation level is recorded over a frequency range and is referred to as the tonemap.
Figure \ref{fig:snr_tm_two_days} shows an example of a tonemap for 192 assessments that correspond to two days.
In general, a higher SNR value and a higher modulation level correspond to a better connectivity in terms of data throughput. 

\subsection{Metadata}
In addition to the SNR assessment multiple metadata information were collected for each PLC modem. 
We distinguish between two types of metadata: cable related and environmental related metadata.

\subsubsection{Cable related metadata}
Cable related metadata covers all information that were available from the cooperating grid operators or collected during the PLC modem installation. \\
Available metadata: \textit{Age, length, number of wires, cross-section, voltage level, number of t-joints, cable type and number of sections} (if a connection between two PLC modems splits up into multiple parts).
Due to missing historical records it was necessary to approximate some very old entries for the number of t-joints and the age.
Oldest cable connections in this dataset were estimated to be installed around 1955.
Approximated values are indicated in the dataset.
See table \ref{tab:metadata} for an overview of all available metadata information. 

\subsubsection{Environmental related metadata}
To investigate environmental influences on the PLC connection quality we provide weather data for each of the locations.
Due to the limited spatial extent of each location (below 800 meters) we assume a homogeneous weather inside each location.
Provided weather data covers in an hourly fashion: temperature, humidity, wind speed, cloud coverage and amount of rain.
A full list of all available weather data fields and for a detailed description we refer to the documentation of OpenWeather \cite{openweatherDocu} were the weather data was acquired. 

\begin{table}
    \centering
    \resizebox{\textwidth}{!}{\begin{tabular}{lr}
        Data field & description \\
        \hline
        year of installation & Year in which the cable/cable section was installed \\
        year approximated & Whether an installation year was approximated or not \\
        cable section & Index of the respective cable section \\
        length & Length of the cable/cable section, in $m$ \\
        number of conductors & Number of conductors \\
        cross section & Conductor cross-section, in $mm^2$ \\
        voltage level & Voltage level, MV = mid voltage; LV = low voltage \\
        t-joints & Number of t-joints within the respective cable section \\
        type &  Type of the respective cable/cable section\\
    \end{tabular}}
    \caption{Summary of the available metadata}
    \label{tab:metadata}
\end{table}

\section{Patterns, Trends in FiN Dataset}\label{sec:patterns}
\begin{figure}[!h]
    \centering
    \includegraphics[width=0.7\textwidth,trim={0cm 0cm 0cm 0cm},clip]{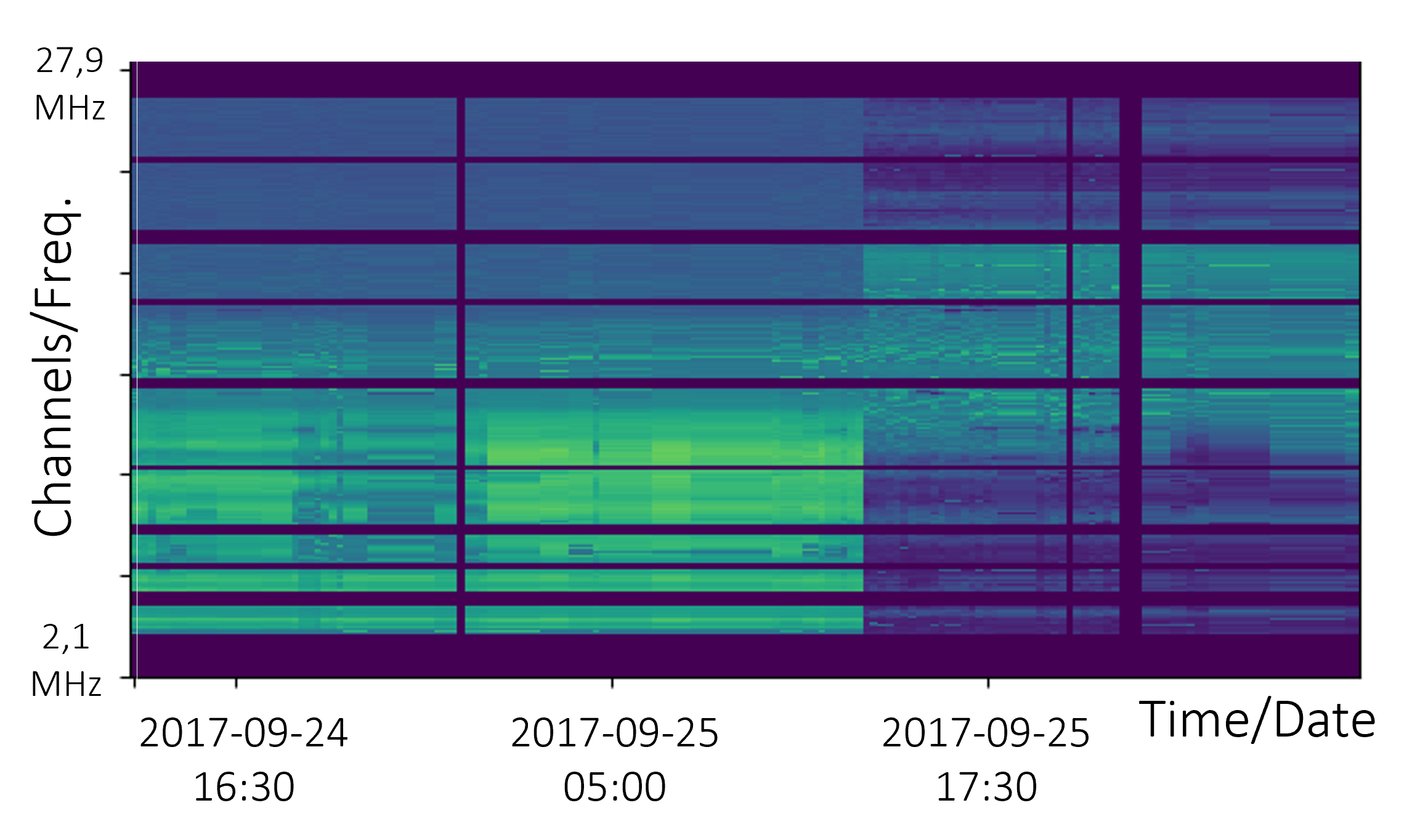}
    \caption{The SNR spectrum shows an imitated fuse failure. On 25th of September the SNR spectrum shows a significant break down due to an imitated fuse failure. The fuse failure was simulated by using a coupling capacitor. The effect in the SNR is similar to a a high-pass filter.}
    \label{fig:fuse_failure1}
\end{figure}

During the full 21 months of collecting data many different trends, events and pattern could be observed. 
In general, all events are classified into different groups based on two characteristics. 
The first is the periodicity, that is, whether an event occurs once or in a periodic scheme.
The second is the time scope of an event, that is, whether an event takes place in a scope of hours, days, months or even if it lasts for ever on.
For example: Figure \ref{fig:fuse_failure1} shows the SNR pattern of a fuse failure that takes place once and has a lasting effect.\\

However, it is not possible to fully outline the reasons of anomalies, as it is not known what effects will be visible in the data.
While the reason for a switching operation in the grid is obvious, most of the sources of interference remain unknown. 
Since even the operators of the electricity grid have no economic means to determine the causes, this information cannot be provided.
Basic physical factors, like signal reflection e.g. at joints, as well as poorly shielded electronics are therefore assumed to be the main cause of the general connection quality and most interference sources.
Contrary to our expectations, however, no significant influence due to the load could be noticed. 
Although there are fluctuations within a day, these do not correlate with the expected standard load profile. 
The following is an overview of different groups of visible behaviours.

\subsubsection{Spiking disturbers \& timeouts}
\begin{figure}[!ht]
    \centering
    \includegraphics[width=0.7\textwidth,trim={0cm 0cm 0cm 0cm},clip]{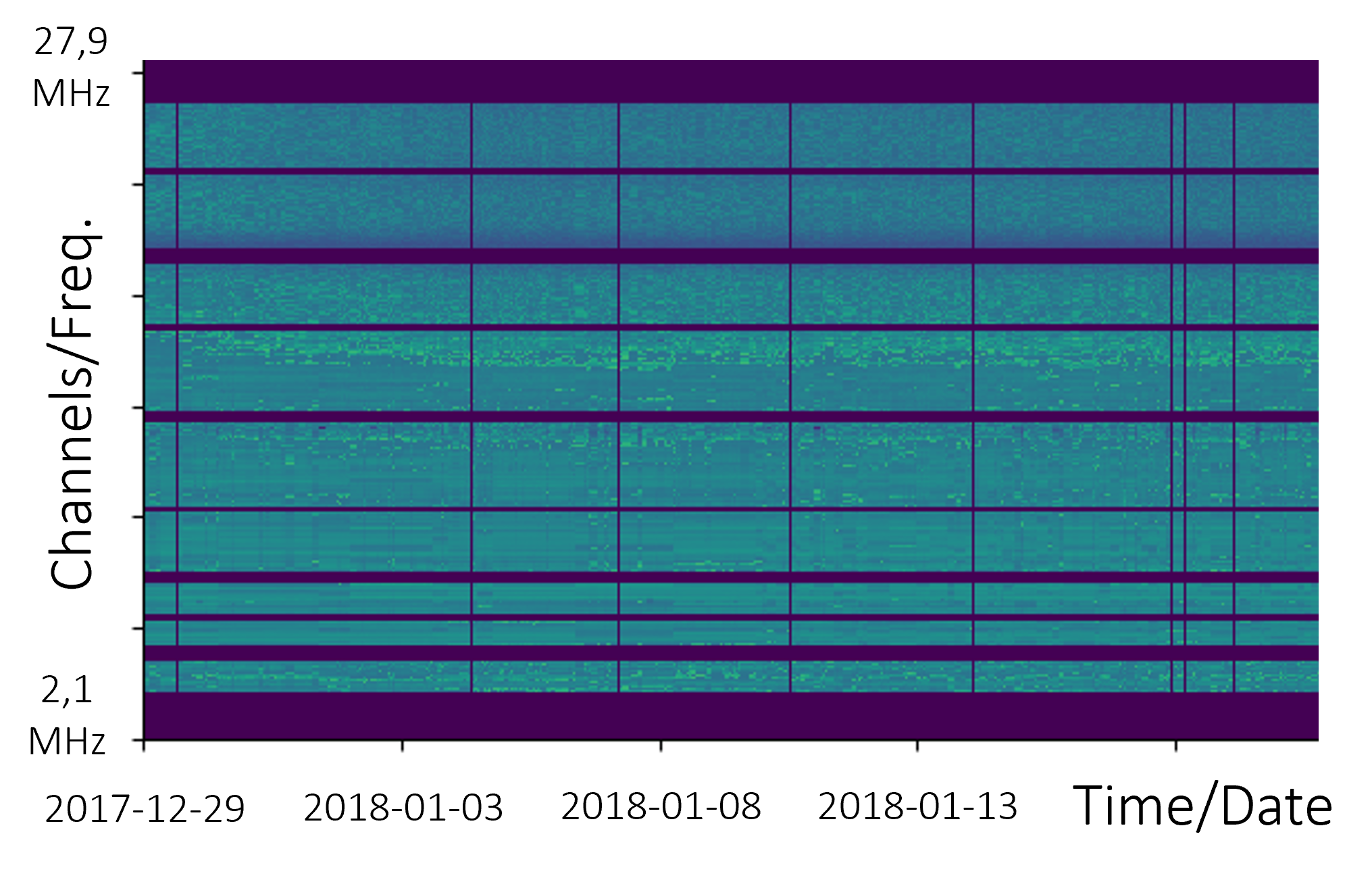}
    \caption{Spiking disturbances and timeouts. The spectrum shows the SNR course over time and channels. Horizontal dead spaces in the spectrum are due to channels or frequencies that are not cleared for use for PLC communications. The noisy course with occasional timeouts corresponds to the normal case for many nodes.}
    \label{fig:spike_noise}
\end{figure}

Spiking disturbances similar to salt and pepper noise, including complete timeouts, are part of the normal behaviour, even of nodes with an overall excellent signal quality. 
But especially in case of timeouts it is not possible to determine the timeout cause due to the absence of background information.
Figure \ref{fig:spike_noise} shows an example of an SNR spectrum that shows spiking disturbances and timeouts. 
\begin{figure}[!ht]
    \centering
    \includegraphics[width=0.7\textwidth,trim={0cm 0cm 0cm 0cm},clip]{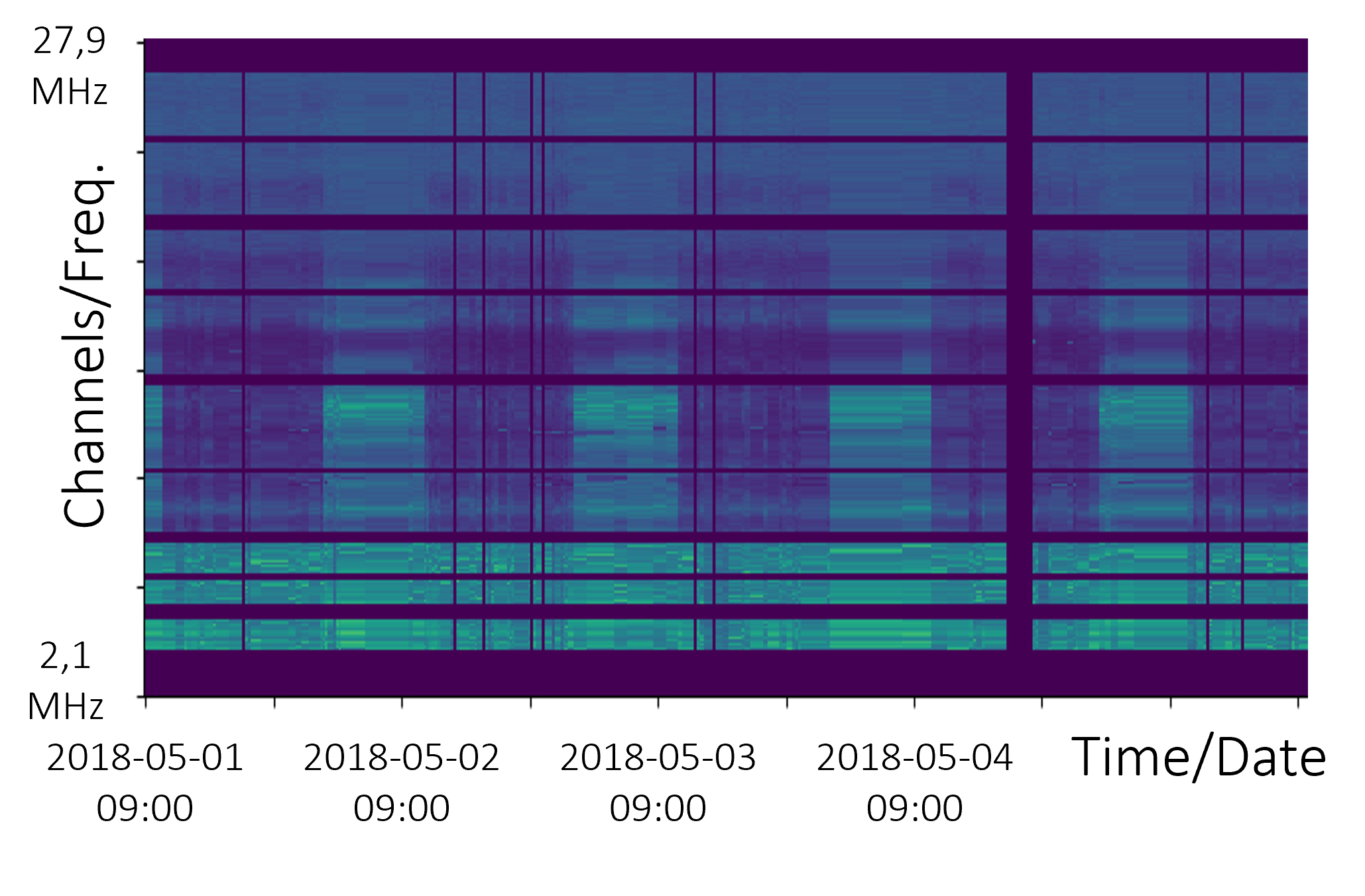}
    \caption{Periodic disturbances in the SNR spectrum of a connection. The time intervals of the disturbances match with the opening hours of an adjacent tanning salon.}
    \label{fig:periodic_dist}
\end{figure}

\subsubsection{Periodic disturbances}
When comparing the SNR profiles over time, a general periodicity becomes visible in most nodes. 
Since the sun and human behaviour are two major impact factors, this periodicity is particularly evident on a daily basis. 
Causes for periodic disturbances could be for example poorly shielded hardware, broken inverters or street lighting.
However, since it is not possible to track down the cause for observed anomalies the actual source stays unknown. 
Figure \ref{fig:periodic_dist} shows an example with a significant SNR breakdown between 09:30am to 11:00pm.
A manual evaluation of the area around the node showed that, the time window of the breakdown matches the opening hours of a neighbouring tanning studio.
Nonetheless, it was not possible to proof this correlation during the project time, but the granularity that has to be dealt with becomes clear. 

\subsubsection{Trends}
\begin{figure*}[!h]
    \centering
    \includegraphics[width=\textwidth,trim={0cm 0cm 0cm 0cm},clip]{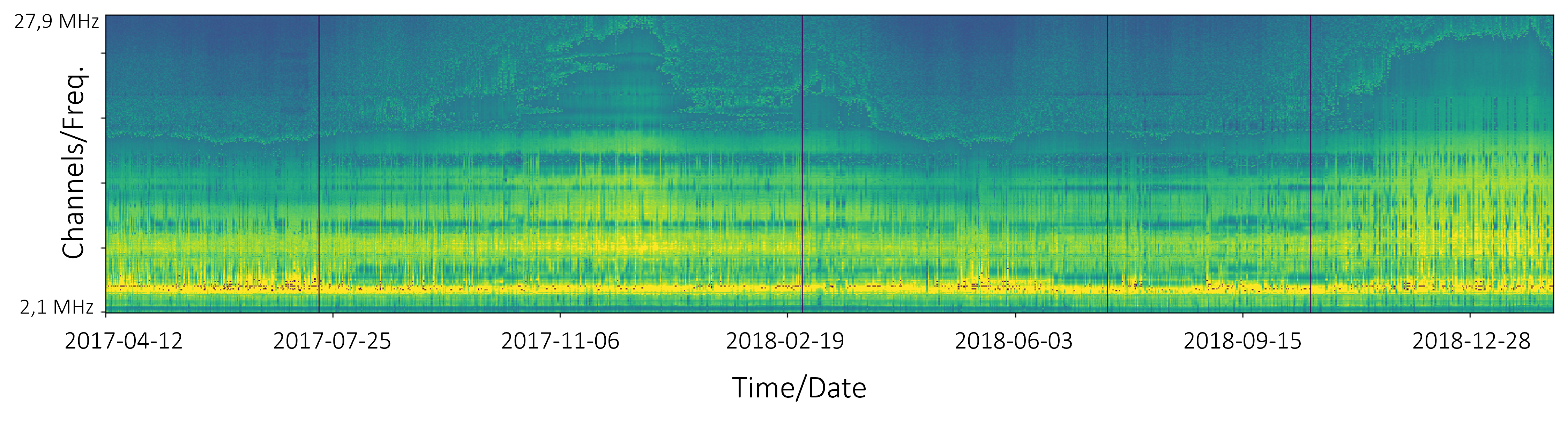}
    \caption{Seasonal change in SNR. Especially high the higher frequencies respectively channels, show a seasonal fluctuation in the SNR. While high frequencies seem to be affected by interference in the summer months, the winter months show better signal quality.}
    \label{fig:trends}
\end{figure*}

Electricity grids are subject to constant change. 
As reported in \cite{hopfer2017new,hopfer2017identification,rezaei2018new}, changing cable properties due to cable age in particular can affect the SNR spectrum and result in trends on a long term perspective.
Furthermore, other reasons for trends are seasons, changing grid topology or changing human behaviour like increasing energy demand.
Figure \ref{fig:trends} shows an example of an SNR spectrum under seasonal change. 

\subsubsection{Misc}
The last group covers all effects that are related to the normal operation of the grid. 
On the one hand side, this could be a switching operation as an expected behaviour, but on the other side also a fuse failure as an unexpected behaviour.
Other examples are cable breakdowns, partial discharges and the installation of new grid hardware like coupling capacitors.
As an example for this group, Figure \ref{fig:fuse_failure1} shows the SNR pattern of a fuse failure imitated using a coupling capacitor. 

\section{Fields of application}\label{sec:apps}
To highlight the novelty and utility of the FiN dataset we want to shed light on different potential fields of application. 
Therefore we present the estimation of the joint count within a cable section in more detail and give in addition an overview of other examples.
Underpinning the field of asset monitoring with this joint count estimation we also want to discuss the grid monitoring in general and security aspects
All experiments are based just on the SNR spectrum and do not involve any onsite measurements. 
We especially encourage the use of Machine Learning (ML) to make use of the vast amount of data that is included in FiN. 
However, this overview is not meant to be limited to the applications on it, but rather to encourage one to explore even more applications on top of it. 
Therefore, FiN is proposed as a new tool to generate new ways to deal with the manifold changes in electricity grids.

\subsection{Asset monitoring}\label{sec:sleeves}
Asset monitoring is an advantage of the wide use of PLC infrastructure.
An example of these assets are the cables themselves, which can contain various numbers of joints.
There are different types of joints, which either connect two different cables (transition joints), two identical cables (connecting joints) or a branch to a third end point e.g. for a house connection (t-joints).
Due to the individual characteristics of the joints, or due to a different quality of installation, the joints on a cable section should have an influence on the SNR spectrum. 
To evaluate this, the FiN dataset provides metadata of the cables.
Based on the number of sections of a cable connection and the cable type, respectively, transition and connection joints can be determined.
The t-joints are also explicitly indicated. 
To estimate the number of joints within a cable section a MLP is used as a baseline, which performs the regression task on the basis of individual SNR profiles. 
As an extended approach, a ResNet18 is utilized, which uses the SNR spectrum of one day as input.

\subsubsection{Data Preparation}
The FiN dataset provides 38 nodes, which are randomly divided into training set with 27 nodes and test set holding 11 nodes.
Since the SNR spectrum is highly correlated in the back and forth directions of two nodes, attention should be paid not to split them between the training and test datasets. 
Timeouts are not considered for training and no further data augmentation is performed. 
Finally, the training batches are scaled between 0 and 1.

\subsubsection{Baseline and ResNet18} 
A simple 3-layered MLP is used as a baseline, which receives a single SNR profile of a connection as input and uses this to determine the number of joints.
Similarly, a ResNet18 is trained, which receives a section of the SNR spectrum as input instead of a single SNR profile. 
The section covers 96 time steps, which corresponds to a time span of 24 hours.
Furthermore, during the generation of the training batches, it is ensured that no timeouts are included in this section.
In both cases, an mean absolute error(MAE) is used as loss, which has the advantage of being less sensitive to outliers.
Over time, the SNR spectrum changes and may even fluctuate periodically based on days, weeks or seasons. 
Since these fluctuations can affect the predicted number of joints, it is primarily important that the mean or median prediction is correct.
Outliers, on the other hand, which are caused by peaks or short-term noise, are therefor not problematic. 
The results are summarised in table \ref{tab:resultsSleeves}.
\begin{figure}[!h]
    \centering
    \includegraphics[width=\textwidth,trim={0.0cm 0.0cm 0cm 0cm},clip]{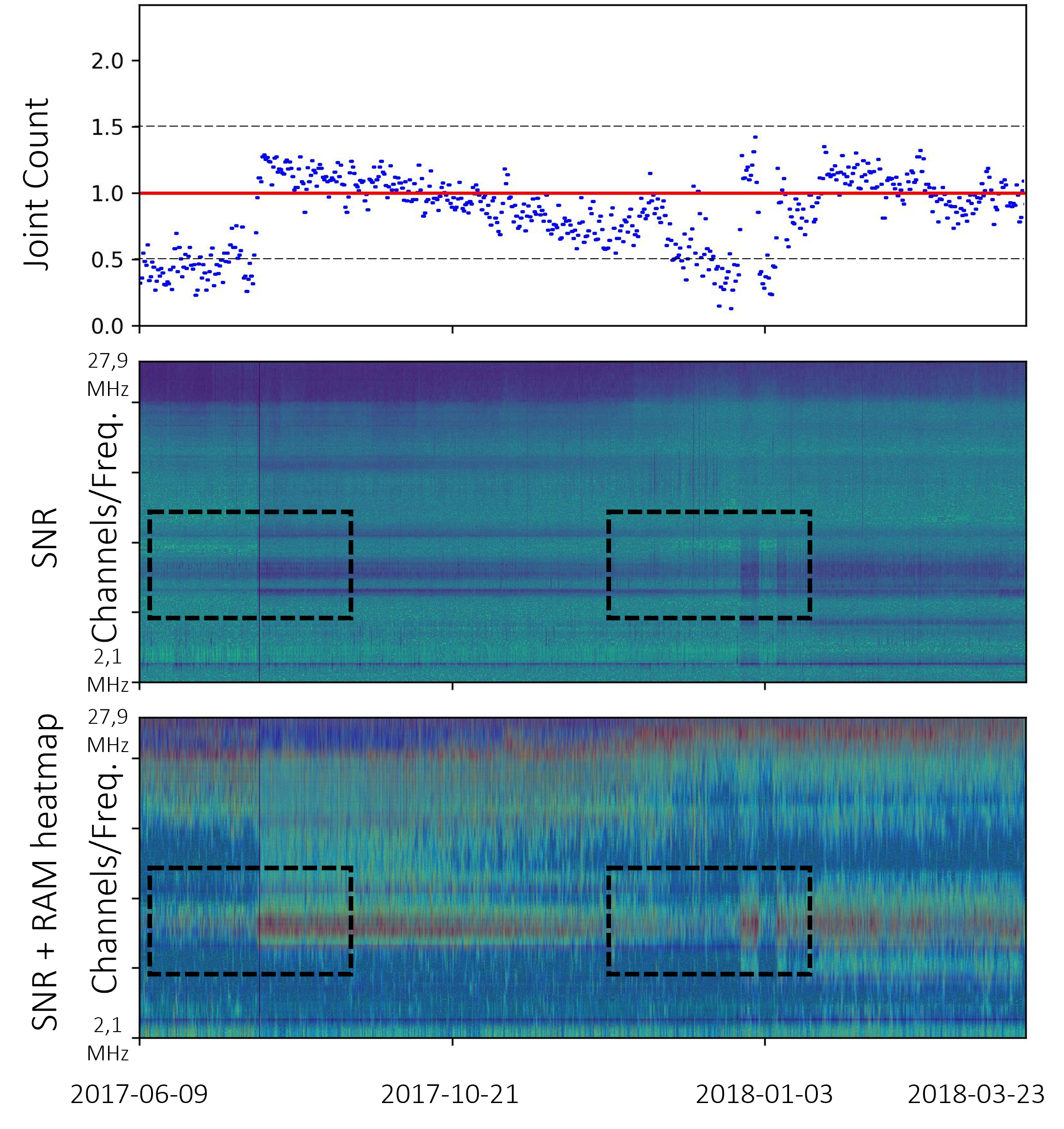}
    \caption{Connection joint estimation results for a joint over time. The left side shows the regression results and the corresponding SNR spectrum, while the right side overlays the regression activation in addition to the same SNR spectrum. Sudden changes in SNR, as indicated by dashed boxes, can significantly affect the regions considered for the regression task.}
    \label{fig:joints_result}
\end{figure}
\begin{figure*}[t!]
    \centering
    \includegraphics[width=\textwidth,trim={0cm 0cm 0cm 0cm},clip]{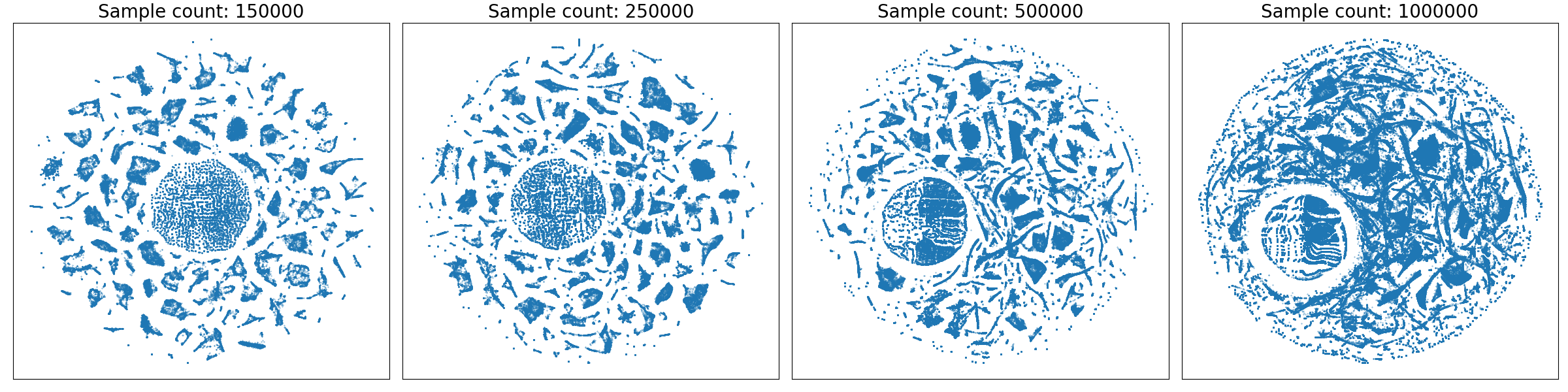}
    \caption{Visualisations of t-SNE embeddings at different samples sizes from the FiN dataset.}
    \label{fig:comparison_TSNE}
\end{figure*}

\subsubsection{Results}
In our experiments, the ResNet18 was able to clearly outperform the MLP architecture. 
Nonetheless, a more detailed analysis of the results shows that, as expected, the accuracy of the results varies significantly. 
However, we assume that providing the SNR spectrum of a whole day as input for the ResNet allows the model to learn features that are more resilient to spiking noise.
Figure \ref{fig:joints_result} shows an example of the spectrum of a connection, the corresponding results from the ResNet, and an overlay that was created using Grad-CAM \cite{selvaraju2017grad}.
The overlay corresponds to the sensitivity in relation to the regression result of the model depending on the PLC channels involved. 
With high sensitivity, the given regions of the spectrum have a strong impact on the regression result. 
It is clearly visible that the regions involved in the sections in which an accurate determination of the number of joints is available deviate strongly from the regions that lead to other results.
In the SNR spectrum it is also clearly visible that the parts for which a poor prediction was made deviate strongly from the remaining parts. 
We assume that the dataset as a whole still contains too few connections and thus variants of joint installations to train a model that can generalise without limitations.
The influence of each joint on the SNR spectrum is expected to vary over time and will also depend on the quality of the installation. 
Since both would require manual inspection on site, we use FiN to mediate this and exploit the results so far as a basis for further investigation into the aspect of how sleeves affect the SNR spectrum over time.

\begin{table}[h]
    \label{tab:resultsSleeves}
    \centering
    \begin{tabular}{l|cccc}
                            & $MAE_{train}$ & $MSE_{train}$ & $MAE_{val}$ & $MSE_{val}$ \\
     \hline
    $MLP_{tj}$                      & $0.49$ & $0.91$ & $4.16$ & $70.97$   \\
    $MLP_{cj}$                      & $0.11$ & $0.08$ & $1.79$ & $7.59$   \\
    $ResNet18_{tj}$                 & $1.33$ & $12.79$ & $2.35$ & $19.69$   \\
    $ResNet18_{cj}$                 & $0.05$ & $0.01$ & $0.82$ & $0.98$   \\
    \end{tabular}
    \caption{Joint estimation results; cj = connecting joints and tj = t-joints}
\end{table}

\subsection{Grid Monitoring}
The ongoing digitalisation and automation of the electricity grid is opening up new ways to use this data for monitoring. 
By using a decentralized infrastructure such as a PLC network, comprehensive monitoring of the electricity grid becomes feasible, even at the low-voltage level. 
Applications within grid monitoring are, for example, the localisation of disturbances and anomalies, the detection of fuse failures and partial discharges.
To tackle these challenges the SNR spectrum was already proofed as a good candidate \cite{huo2021power,bondorf2021broadband}.
However, we want to address the aspect of grid monitoring from a different direction. 
By using the complete FiN dataset to find a suitable low dimensional representation and subsequently clusters within this representation, we aim on defining clusters of similar connection states.
The SNR spectrum shows overall rich information concerning many different influences, however it also shows a lot of noise that does not affect the general state of a PLC connection.
By merging similar SNR profiles into the same cluster we want to support interpretability of the SNR spectrum and provide a good foundation for downstream tasks like anomaly detection. 
In the following, we show how we built a low dimensional representation of the SNR data and how we obtained cluster labels.
\begin{figure}[b!]
    \centering
    \includegraphics[width=0.6\textwidth,trim={17cm 0cm 7.3cm 9cm},clip]{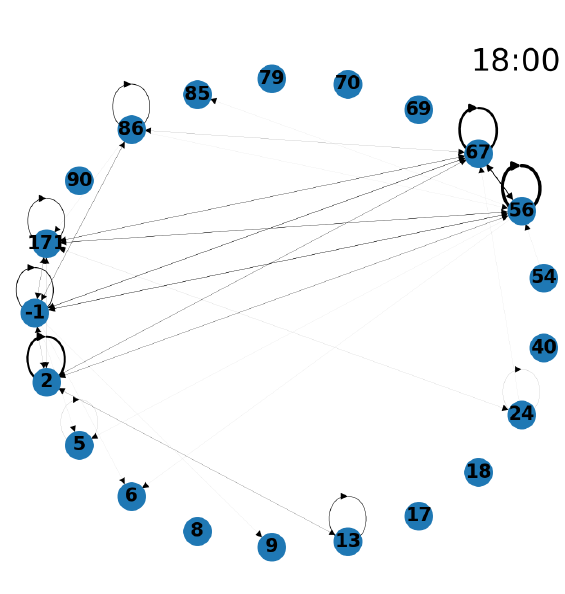}
    \caption{Transition graph at 6am. Based on the past sequence of visited states, the transition graph illustrates the chance to switch from one state to another. As can be well illustrated in the figure \ref{fig:circular}, the probability of being in a certain state changes over the course of a day. The strength of a connection indicates the probability of the specified transition.}
    \label{fig:graph_day}
\end{figure}

\subsubsection{Low-dimensional Representation}
A low-dimensional representation, especially in 2 dimensions, support computational efficiency as well as providing a good interpretability in terms of visualization. 
While the former is important when thinking about a electricity grid that consists of many thousands of nodes that need to be monitored, the latter aspect is important for people during a manual inspection.
In order to build a low-dimensional representation, t-SNE \cite{van2008visualizing}, UMAP \cite{mcinnes2018umap} and MDS \cite{kruskal1964nonmetric} were utilized and within the experiments it could be shown that all approaches have their limitations.
While t-SNE and UMAP show good results when the number of samples is kept low (e.g., 150,000 samples), both approaches tend to break down when a million or even the entire data set is used. 
Figure \ref{fig:comparison_TSNE} shows an example of different low-dimensional representations using t-SNE when using different sample sizes. 
In contrast to t-SNE and UMAP, the MDS approach shows significant drawbacks due to its memory complexity of $O(N^2)$, wherefore MDS was not used for any further experiment.
Due to these limitations, more sophisticated approaches such as machine learning based (\cite{tian2017deepcluster,yan2020clusterfit}) techniques are suggested.
Nevertheless, the next two sections discuss how a low-dimensional representation, as shown in Figure \ref{fig:comparison_TSNE}, can be used to detect anomalies and monitor the connection states within the PLC network.

\begin{figure*}[t!]
    \centering
    \includegraphics[width=\textwidth,trim={0cm 0cm 0cm 0cm},clip]{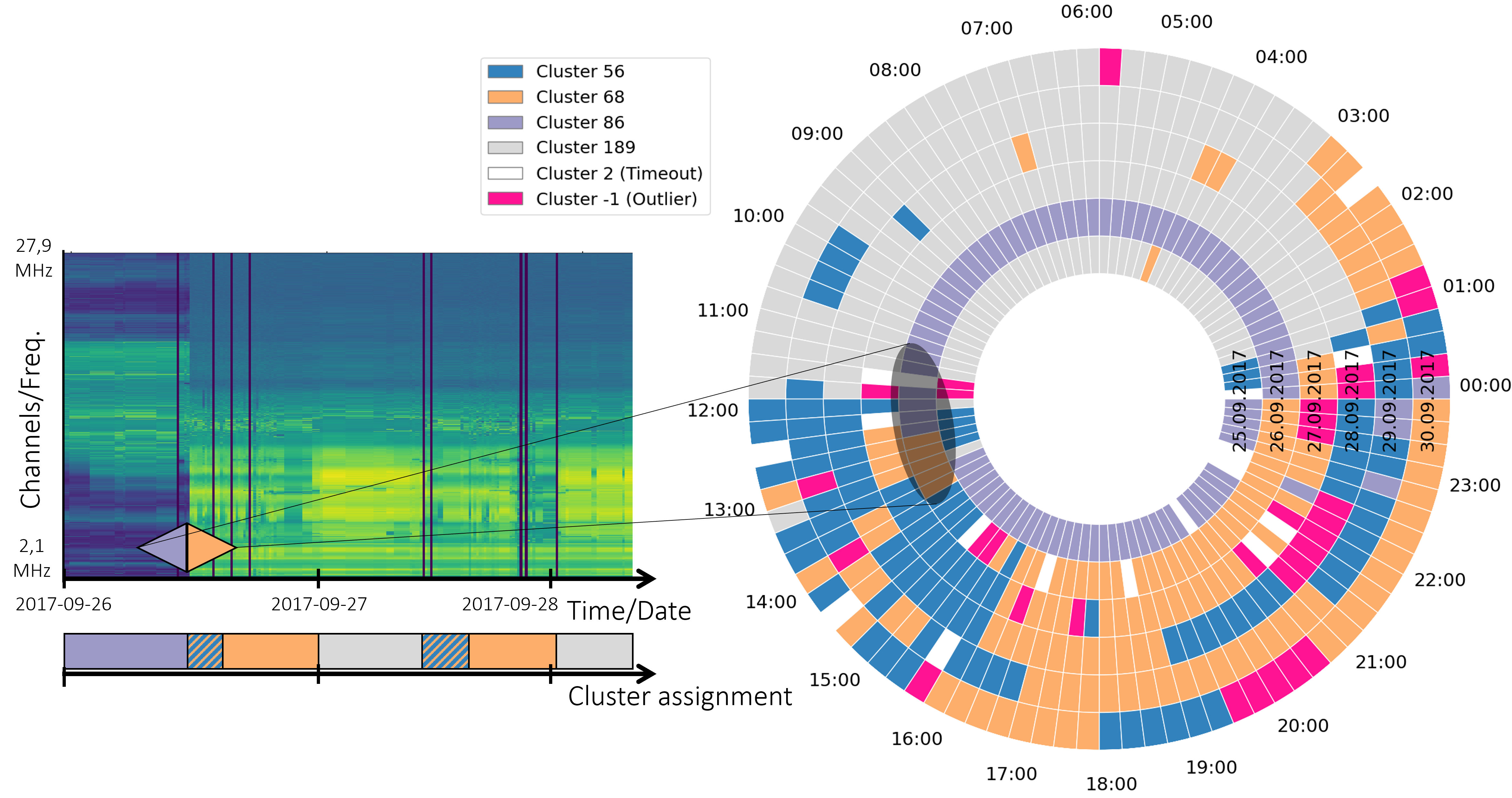}
    \caption{A excerpt from the SNR spectrum (left) and the corresponding concentric visualisation of cluster assignments (right). The time range shows an imitated fuse failure. Two triangles show the SNR spectrum before (purple) and after (orange) uninstalling a coupling capacitor used to simulate a fuse failure. The corresponding area in the concentric diagram is highlighted as well. Overall, the concentric arrangement of the diagram provides good access to insights about the data. At the same time, it is possible to distinguish at first glance between random timeouts and time-dependent trends.}
    \label{fig:circular}
\end{figure*}
\subsubsection{Clustering}
Providing low-dimensional data is not only valuable for downstream machine learning tasks, but can also be used to distill information from the data.
When operating a electricity grid, decisions (e.g. scheduling a, on site inspection) have to be made on the basis of grid states and recognition of issues, which is why a comprehensible representation of the grid state is essential.
One way to provide such a representation is a comprehensive visualization.
To construct a visualization based on the raw SNR spectrum that can also be read by non-specialists, t-SNE is applied to a subset of 250,000 samples. 
Afterwards, DBSCAN is utilized to obtain labels for the low-dimensional data. 
In the following, the cluster assignments are used for the visualization instead of the SNR profiles.
Figure \ref{fig:circular} shows a visualization of the cluster assignments in form of a radial chart. 
Since the course of the state of a PLC connection is strongly dependent on human processes, a visualisation on a 24-hour basis is appropriate.
By such a concentric arrangement, noise can be clearly distinguished from a trend, and changes that extend over several days can be easily recognized. 

\subsubsection{Anomaly discovery}
Anomaly discovery in SNR spectra is one of the most challenging applications in this field.
Even when an anomaly is reliably discovered, the cause of the anomaly remains unknown, as it is usually very costly to find out. 
Therefore, because the origins are largely unknown and the anomalies usually have no direct impact on grid operations, it is not possible to determine anomalies of interest. 
In addition, a vast number of PLC connections show significant fluctuations that at first glance appear to be an anomaly.
It exists a wide range of influences, which act as anomalies in the SNR spectrum, but have no influence on the electricity grid or the power line communication. 
On the other hand, the anomalies carry a wide range of potential information.
Figure \ref{fig:circular} shows a fuse failure which could be within the SNR spectrum as well as in the circular visualization. 
Since, connection states are described by discrete cluster labels we suggest approaches like Hidden Markov Models (HMM) \cite{eddy2004hidden}, Dynamic Time Warping \cite{muller2007dynamic} or Temporal Convolutional Networks \cite{oord2016wavenet} to detect anomalies.
To provide a first glance at this topic, Figure \ref{fig:graph_day} shows a transition graph representing the probability that a given PLC connection switches between two states at a given time.
This graph could be used directly to calculate a certainty for the current state of being an anomaly or to estimate the same based on the last hours.
Beyond that, nevertheless, the use of more sophisticated probabilistic approaches is recommended. 
However, this aspect is subject to further research and is briefly discussed in section \ref{sec:future_work}.

\subsection{Security}
We also want to shed light on the idea of using a PLC-based infrastructure as additional security layer, to monitor the electricity grid. 
Due to their distributed, independent and flexible nature, PLC systems are well suited to act as a redundant control mechanism to check against traditional grid monitoring.
Since all nodes act independently, the network is hard to compromise and even in the event of single failures, the network is able to adapt if alternative routes are available. 
Furthermore, since each cable connection is measured from both sides (transmitting and receiving), adjacent nodes can check a signal for plausibility.
Because of these properties, PLC-based monitoring systems might be a useful tool for the detection of attacks against critical electricity infrastructure. 

\section{Conclusion}
In this paper, we proposed and analysed a new real data set collected during practical operation of the power system.
Our work emphasizes the high value of the data in a variety of potential application domains, as well as the richness of its nature, consisting of a mixture of structured, dynamic and static information.
The SNR data presented here open up applications that go beyond what previous datasets based on consumption and voltage values allow.
The applications shown are intended to give a first impression of the wide range of possible applications of the dataset.
Furthermore, we were able to successfully show how the SNR data can be used to estimate the joint count within a cable section.
The key advantage of our approach is that no additional knowledge about the infrastructure (cable length, type, etc.) is necessary. 
Only the SNR spectrum was used for this task. 
We believe that further research on this topic will yield a variety of new research challenges as well as important developments in PLC-based monitoring of smart grids.
The final section \ref{sec:future_work} mentions further potential application possibilities which will be pursued in the future.

\section{Data Availability}\label{sec:dataAvailabilty}
FiN dataset is available under Creative Commons CC BY 4.0 International and can be downloaded from Zenodo \cite{FiNDataset}.

\section{Future Work}\label{sec:future_work} 
The digitalisation of electricity grids is currently picking up speed and is coupled with significant opportunities, risks and challenges that still need to be overcome.
Possibilities and use cases that we have addressed so far can only shed light on a small aspect.

As mentioned earlier, the FiN dataset represents a novel data set that can be used to address various challenges arising from the transition to a smart grid.
Therefore, we set out to take a first glance at what is hidden in this previously widely unrecognized SNR data from PLC networks.
However, the FiN datasets are only a first step in this area and represent a proof of concept in several directions. 
To advance this new area of SNR-based network assessment, we intend to address several open questions in our future research:

\begin{itemize}
    \item Pholovoltaic systems: \cite{lopez2017noise} suggests that noise from PV inverters can be detected within PLC networks, consequently it could be possible to find a correlation between noise in SNR and weather. Future work should further investigate these correlations and prove them with practical examples. In addition, an automatic detection of new PV systems connected to the grid will also be evaluated.  
    
    \item Predictive maintenance: Encouraged by our results, we conclude that the SNR spectrum contains rich information about a wide range of cable properties.
    Challenges to be investigated in the future are, for example, to what extent AI methods are suitable for predicting the remaining lifetime of cable installations or the extent to which a grid state identification can be implemented. 
    
    \item Cable joints: During the collection of the data, it was noticed that a significant proportion of failures result in particular from failing joints. Further analysis of the joints and their PLC spectrum is therefore subject of further research. 
    
    \item Topology estimation: As discussed in section \ref{sec:in_direct_neighbours} PLC cannot provide information on the type of neighborhood, whether an adjacent node is a direct or indirect neighbor. Therefore, we propose the SNR spectrum as a carrier for fingerprints of specific cable sections to determine whether two neighbors are directly or indirectly adjacent. Future research will address AI-based automatic detection of PLC network topology. 
    
    \item Privacy and cyber-security: The risks associated with any digitalisation are software bugs and vulnerability to attacks. For this purpose, findings from measurements of the PLC infrastructure could also be used to plausibilise measurement data and network states of the legacy grid control systems on a second, independent level. In this regard, questions also arise concerning the privacy of customers who are forced to connect to such a system during a PLC rollout. Even if full-scale expansion is still a long way off, we would like to encourage discussion of these issues. 
    
    \item Additional measurements: Since PLC networks provide a good platform to collect various measurements from the nodes, future research will address the extent to which other measurements, such as voltage, can contribute to more comprehensive predictions regarding the application areas presented.
    
    \item Robustness: Due to the comparatively small spatial extent of the area, we want to explore the possibility of transferring the models and findings obtained in FiN to other grid areas. These include, above all, the future research topics already presented, as well as the approaches shown in this work.
\end{itemize}

To address all of the future work mentioned above, a second dataset is currently being created that uses the previous results as a proof-of-concept and, in particular, increases the number of nodes involved to several thousand. 
The massively larger resulting dataset is expected to yield more robust results in all directions.  

\section{Acknowledgment}
The presented work in this publication is based on research activities, supported by the Federal Ministry of Education and Research (BMBF) and the Projektträger Jülich (PTJ), the described topics are included in the project “Fühler im Netz 2.0” (reference number: 03EK3540B).
Only the authors are responsible for the content of this paper.

\begin{figure}[!h]
   \begin{minipage}[!h]{.3\linewidth} 
      \includegraphics[width=\linewidth]{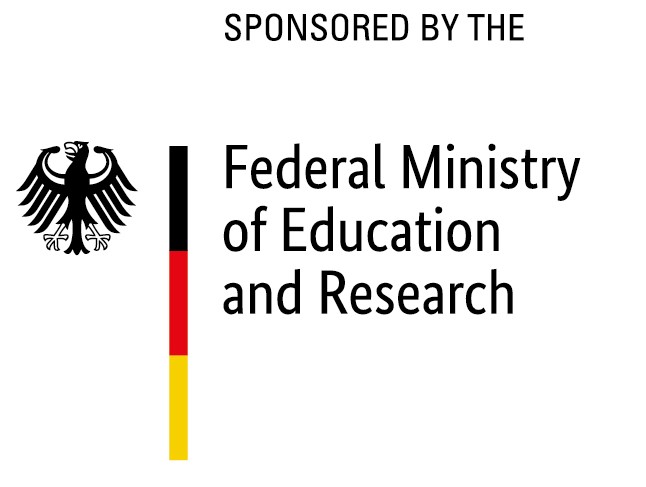}
   \end{minipage}
   \hspace{.1\linewidth}
   \begin{minipage}[!h]{.3\linewidth} 
      \includegraphics[width=\linewidth]{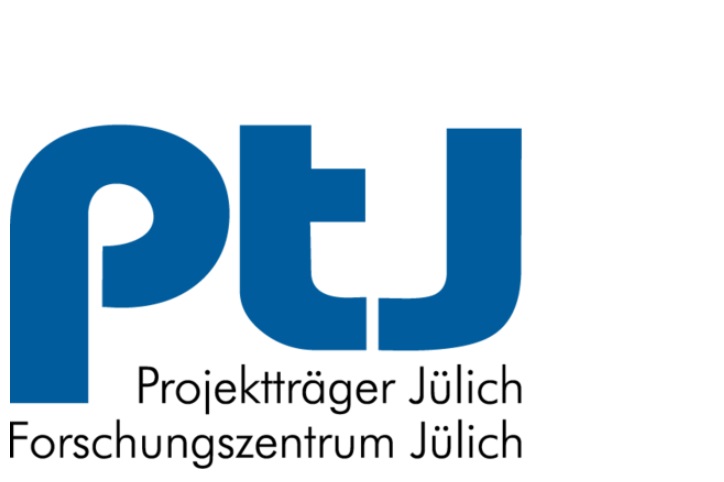}
   \end{minipage}
\end{figure}

\appendix

\bibliographystyle{elsarticle-num} 
\bibliography{cas-refs}





\end{document}